\documentclass[fleqn,usenatbib]{mnras}

\usepackage{newtxtext,newtxmath}

\usepackage[T1]{fontenc}
\usepackage{ae,aecompl}
\usepackage{soul}

\usepackage{float}

\usepackage{graphicx}	
\usepackage{amsmath}	
\usepackage{amssymb}	

\newcommand{\hMpc}{{\textrm{ $h^{-1}$Mpc}}}
\newcommand{\kms}{{\textrm{ km\,s$^{-1}$}}}

\newcommand{\hMsun}{{\textrm{ $h^{-1}$M$_{\sun}$}}}
\newcommand{\Wh}{\mathrm{W}}
\newcommand{\Lh}{\mathrm{L}}
\newcommand{\Mh}{\mathrm{M}}
\newcommand{\winner}{{\emph{winner}}}
\newcommand{\loser}{{\emph{loser}}}
\newcommand{\median}{{\emph{median}}}

\title[Environment dependences on angular momentum growth]{Deviations from tidal torque theory: environment dependences on halo angular momentum growth}

\author[L\'opez, Merch\'an \& Paz]{
Pablo L\'opez,$^{1,3}$\thanks{E-mail: plopez@oac.unc.edu.ar}
Manuel E. Merch\'an,$^{1,2,3}$
Dante J. Paz$^{1,2,3}$
\\
$^{1}$Observatorio Astron\'omico de C\'ordoba, Universidad Nacional de C\'ordoba (UNC), Francisco N. Laprida 854, C\'ordoba, Argentina\\
$^{2}$Consejo Nacional de Investigaciones Cient\'ificas y T\'ecnicas (CONICET), Rivadavia 1917, Buenos Aires, Argentina.\\
$^{3}$Instituto de Astronom\'ia Te\'orica y Experimental, CONICET-UNC, Laprida 922, C\'ordoba, Argentina\\
}

\date{Accepted XXX. Received YYY; in original form ZZZ}

\pubyear{2018}

\begin{document}
\label{firstpage}
\pagerange{\pageref{firstpage}--\pageref{lastpage}}
\maketitle

\begin{abstract}
The tidal torque theory (TTT) relates the origin and evolution of angular momentum with the environment in which dark matter (DM) haloes form. The deviations introduced by late non-linearities are commonly thought as noise in the model. In this work, we analyze a cosmological simulation looking for systematics on these deviations, finding that the classification of DM haloes according to their angular momentum growth results in samples with different internal alignment, spin parameter distribution and assembly history. Based on this classification, we obtain that low mass haloes are embedded in denser environments if they have acquired angular momentum below the TTT expectations ($\Lh$ haloes), whereas at high masses enhanced clustering is typically associated with higher angular momentum growths ($\Wh$ haloes). Additionally, we find that the low mass signal has a weak dependence on the direction, whereas the high mass signal is entirely due to the structure perpendicular to the angular momentum.
Finally, we study the anisotropy of the matter distribution around haloes as a function of their mass. We find that the angular momentum direction of $\Wh$ ($\Lh$) haloes remains statistically perpendicular (parallel) to the surrounding structure across the mass range $11<\mathrm{log}(M/h^{-1}\mathrm{M}_{\sun})<14$, whereas haloes following TTT show a ``spin flip'' mass consistent with previously reported values ($\sim5\times10^{12}\hMsun$). Hence, whether the spin flip mass of the deviated samples is highly shifted or straightly undefined, our results indicate that is remarkably connected to the haloes angular momentum growth.
\end{abstract}




\section{Introduction}

Dark matter haloes (DM) are the largest virialized structures in the Universe. Their internal dynamics are linked to the history of galaxies, groups and clusters that form {in} their potential wells, as well as to the large-scale structure (LSS) of the Universe, which not only establishes characteristic directions for the matter infall, but also provides an anisotropic tidal field which could affect the DM halo internal velocity distribution. Therefore, understanding the dynamical properties of haloes becomes a keystone in any cosmological and galaxy formation model.

A first attempt to perform a detailed calculation on the origin of the rotation of protogalaxies was carried out by \citet{peebles1969}, {who used first-order approximations to the density and velocity fields to estimate} the angular momentum growth of the matter contained in a comoving spherical region of a expanding universe. His results indicated that the acquisition of angular momentum is a second order effect. \citet{doroshkevich1970} showed, however, that the angular momentum of a protogalaxy grows at first order, and that the conclusion reached by \citeauthor{peebles1969} was a consequence of the spherical symmetry he imposed to solve the equations. Although \citeauthor{doroshkevich1970} did not provide many details about his calculations, in a later work, \citet{white1984} made a more detailed development of the model, improving its predictions and limitations. 

These studies gave rise to the most accepted model for angular momentum acquisition in the current paradigm, the tidal torque theory (TTT), which basically states that most of the angular momentum of haloes is gained during the linear stage of structure formation, due to the misalignment between its shape tensor and the tidal shear field exerted by the surrounding matter distribution. The decoupling of the protohalo region from the general expansion of the Universe, when it turns around to non-linear collapse and virialization, makes the process lose efficiency. This occurs due to the reduction of the system inertia and the increasing separation with the neighboring matter responsible for the tidal torque. In consequence, {in the context of TTT,} little angular momentum is expected to be tidally exchanged between haloes after the turn around point \citep{porcianietal2002a}. {However, it is to expect that further non-linear interactions will produce deviations from the angular momentum predicted by the model.}

The TTT provides an analytical expression (see section \ref{ttt_formalism}) that allows to compare results in numerical simulations with theoretical predictions \citep{white1984,hoffman1986,barnesyefstathiou1987,heavensypeacock1988,sugermanetal2000,leeypen2000,porcianietal2002a}. Along with this, the model naturally relates the origin and evolution of angular momentum to the characteristics of the environment in which DM haloes form. This is a powerful tool to understand the alignments between the large-scale structure and the shapes and angular momentum directions of DM haloes in numerical simulations \citep{bailinysteinmetz2005,aragoncalvoetal2007,hahnetal2007,libeskindetal2012,foreroromeroetal2014} and galaxies in observational catalogs \citep{leeyerdogdu2007,pazetal2008,tempelylibeskind2013,zhangetal2015}. Nevertheless, it has been shown that the TTT accurately predicts the evolution of angular momentum only during the linear and quasi-linear stages of structure formation \citep{sugermanetal2000,porcianietal2002a}; late stages tend to erase the TTT effects and the angular momentum of DM haloes and galaxies starts to be greatly affected by non-linear processes such as mergers \citep{vitvitskaetal2002,bettyfrenk2016} and the emergence of vortical flow fields \citep{libeskindetal2012}, although it has been suggested that the latter can be reconciled with the TTT if {certain assumptions are made} to account for the larger scale anisotropic environments \citep{codisetal2015,laigleetal2015}.

There is agreement on that the amount of angular momentum acquired by tidal torque is small. DM haloes are supported mainly by velocity dispersion and only marginally by rotation. The present-day distribution of the dimensionless spin parameter $\lambda$, which measures the fraction of coherent rotation in a system compared to random motions, has been computed in numerical simulations and has a well known log-normal distribution that peaks at $\lambda_{0}\sim 0.04$ \citep{bullocketal2001,bailinysteinmetz2005,bettetal2007}, with a very low dependence on the halo masses \citep{vitvitskaetal2002}. {The rotational support of DM haloes, however, has been found to correlate with their environment, with high-spin haloes showing enhanced clustering} \citep{gaoywhite2007,faltenbacherywhite2010}{.}

It has been also long known that the direction of the angular momentum correlates with the shape of the haloes, although some scatter has been detected in these relations. In early CDM low-resolution studies \citep{barnesyefstathiou1987,warrenetal1992} and, more recently, in high-resolution $\Lambda$CDM numerical simulations \citep{bailinysteinmetz2005}, it has been shown that the angular momentum of DM haloes is, at the present time, more frequently aligned with their minor axis and perpendicularly oriented with respect to their major axis. 
Consistently, {the one-halo term of anisotropic correlation functions has shown that the preferred direction of the mass distribution inside DM haloes tends to be perpendicular to their angular momentum} \citep{pazetal2008}.
However, the evolution of the angular momentum direction, and how it relates to other dynamical properties of DM haloes, continues to be poorly explained by the TTT, with studies showing that its implementation predicts angular momentum directions that differ on {$\sim 50\degr$} from the measured values in numerical simulations \citep{porcianietal2002a}.

The scatter, both in direction and magnitude, of the TTT predictions for the present time angular momentum, together with the conflicting results obtained for the angular momentum alignment with respect to the LSS, suggest that, in order to capture the non-linear effects involved in the post-TTT angular momentum acquisition process, a different approach is needed. In this paper we study systematic deviations from the TTT predictions and, considering this, we classify DM haloes according to their angular momentum growth. The resulting halo samples are then analyzed and compared in regard of their dynamical properties and, specially, of the clustering and preferred orientation of their surrounding LSS. 

This paper is organized as follows: in Section \ref{ttt_formalism}, we briefly present elements of the TTT formalism, in order to establish the bases to identify systematic deviations from the model that allows us to build a relevant classification; in Section \ref{metodos}, we present the simulated data and the methods we use to implement the TTT and to define our classification according to the haloes angular momentum growth; in Section \ref{results}, we show the main results of our study; finally, in Section \ref{conclusions}, we summarize our work and discuss some of the implications suggested by our results.

\section{The tidal torque theory formalism}
\label{ttt_formalism}

The TTT predicts the evolution of the angular momentum of a protohalo within the framework of a Friedmann-Lema\^{i}tre-Robertson-Walker universe, assuming that matter is a non-collisional fluid which interacts only gravitationally. During the linear regime, protohaloes are assumed to be small perturbations above the mean universal density, which grow due to gravitational instability while the universe expands according to the scale factor $a(t)$. In terms of the {Lagrangian} coordinates $\mathbfit{q}$, defined as the {comoving} coordinates $\textbf{\emph{x}}=\mathbfit{r}/a(t)$ at $t\rightarrow 0$, where $\mathbfit{r}$ represents the physical coordinates, the angular momentum of a protohalo whose matter is enclosed in a {Lagrangian} volume $V_\mathrm{L}$ can be expressed as:
\begin{equation}
\label{ttt_0}
\mathbfit{J}(t) = \int_{V_\mathrm{L}}[\mathbfit{r}(\mathbfit{q},t)-\mathbfit{r}_\mathrm{cm}(t)] \bmath{\times} \mathbfit{v}(\mathbfit{q},t)\bar{\rho}(t)a^3(t)d^3q,
\end{equation}
where $\bar{\rho}(t)$ is the mean universal density, $\mathbfit{r}_\mathrm{cm}(t)$ is the usual centre-of-mass position at time $t$ and $\mathbfit{r}(\mathbfit{q},t)$ and $\mathbfit{v}(\mathbfit{q},t)$ describe, respectively, the time evolution of the physical coordinates and the speed of the mass element $\bar{\rho}(t)d^3q$ relative to the halo centre. In the comoving coordinates, expression \eqref{ttt_0} can be written exactly as:
\begin{equation}
\label{ttt_1}
\mathbfit{J}(t) = a^2(t)\int_{V_\mathrm{L}}[\textbf{\emph{x}}(\mathbfit{q},t)-\textbf{\emph{x}}_\mathrm{cm}(t)] \bmath{\times} \dot{\mathbfit{x}}(\mathbfit{q},t)\bar{\rho}(t)a^3(t)d^3q,
\end{equation}
where the dot denotes a derivative with respect to the cosmic time $t$. 

The Zel'dovich approximation \citep{zeldovich1970} states that in the linear regime the {comoving} position of a mass element can be well described in time as its {Lagrangian} position $\mathbfit{q}$ plus a displacement {$S(\mathbfit{q},t)$} with spatial and temporal dependence decoupled, i.e. {$S(\mathbfit{q},t)=\mathbfit{f}(\mathbfit{q})D(t)$}. Moreover, the proportionality between the velocity field and the gravitational potential $\psi$, that holds in the linear regime, implies that the spatial dependence of {$S(\mathbfit{q},t)$} is due only to the initial surrounding matter distribution. Thus, we can write the comoving coordinates of a mass element in the linear approximation as $\mathbfit{x}(\mathbfit{q},t)=D(t)\nabla\psi(\mathbfit{q})+\mathbfit{q}$, where $D(t)$ is a function that modules the time evolution according to the growing mode of the density fluctuations. It can be seen that the time derivative $\dot{\mathbfit{x}}(\mathbfit{q},t)=\dot{D}(t)\nabla\psi(\mathbfit{q})$ has the same direction as the displacement $\mathbfit{x}(\mathbfit{q},t)-\mathbfit{q}=D(t)\nabla\psi(\mathbfit{q})$, so the term $\mathbfit{x}(\mathbfit{q},t) \bmath{\times} \dot{\mathbfit{x}}(\mathbfit{q},t)$ in equation \eqref{ttt_1} vanishes. By removing this and replacing $\dot{\mathbfit{x}}(\mathbfit{q},t)$ we obtain:
\begin{equation}
\label{ttt_2}
\mathbfit{J}(t) = a^2(t)\dot{D}(t)\int_{V_L}[\mathbfit{q}-\mathbfit{q}_\mathrm{cm}] \bmath{\times} \nabla\psi(\mathbfit{q})\bar{\rho}(t)a^3(t)d^3q.
\end{equation}
The meaning of the last expression can be clarified a little further by assuming that the scalar potential $\psi$ can be reasonably approximated by its Taylor series around $\mathbfit{q}_\mathrm{cm}$ to the second order:
\begin{equation}
\begin{split}
\psi(\mathbfit{q})= \psi(\mathbfit{q}_\mathrm{cm})+(q_i-q^\mathrm{cm}_i&)\frac{\upartial\psi}{\upartial q_i} \Big|_{\mathbfit{q}_\mathrm{cm}}+\\
&\frac{1}{2}(q_i-q^\mathrm{cm}_i)(q_j-q^\mathrm{cm}_j)\frac{\upartial^2\psi}{\upartial q_i\upartial q_j}\Big|_{\mathbfit{q}_\mathrm{cm}}.
\end{split}
\end{equation}
Replacing in \eqref{ttt_2} and keeping only the terms up to second order, the $i$-th component of the angular momentum $\mathbfit{J}(t)$ can be written as:
\begin{equation}
\label{ttt_3}
\begin{split}
J_i(t) = a^2(t)\dot{D}(t)\epsilon_{ijk}&\frac{\upartial^2\psi}{\upartial q_j\upartial q_l}\Big|_{\mathbfit{q}_\mathrm{cm}}\bmath{\times}\\
&\int_{V_\mathrm{L}}[q_l-q^\mathrm{cm}_l][q_k-q^\mathrm{cm}_k]\bar{\rho}(t)a^3(t)d^3q,
\end{split}
\end{equation}
where $\epsilon_{ijk}$ is the fully antisymmetric rank-three tensor. Despite its unfriendly appearance, equation \eqref{ttt_3} contains known quantities. The integral defines the $lk$-th component of the inertia tensor $\mathbfss{I}$ of the matter contained in $V_\mathrm{L}$, while $\frac{\upartial^2 \psi}{\upartial q_j\upartial q_l} \Big|_{\mathbfit{q}_\mathrm{cm}}$ is the $jl$-th component of the Hessian $\mathbfss{T}$ of the gravitational potential around {the centre of mass of the protohalo}\footnote{It can be seen that only the shape tensor and tidal tensor, i.e. the detraced parts of $\mathbfss{I}$ and $\mathbfss{T}$, respectively, contribute to the cross product in equation \eqref{ttt_3}.}. With these definitions we obtain the basic expression of the TTT \citep{white1984}:
\begin{equation}
\label{ttt_equation}
J_i(t) = a^2(t)\dot{D}(t)\epsilon_{ijk}T_{jl}I_{lk}.
\end{equation}

Equation \eqref{ttt_equation} expresses the fact that, in the TTT framework, the components of the angular momentum grow due to the cross product of the inertia tensor $\mathbfss{I}$, which only depends on the distribution of the matter within the protohalo, and the Hessian $\mathbfss{T}$, built from partial derivatives, which makes it dependent on the distribution of neighboring density perturbations. Consequently, the torque depends on the protohalo shape, the external tidal field and the {misalignment} between them \citep{porcianietal2002a}, and it can be time evolved, according to the Zel'dovich approximation, by the factor $a^2(t)\dot{D}(t)$. 

\begin{figure}
\includegraphics[width=\columnwidth]{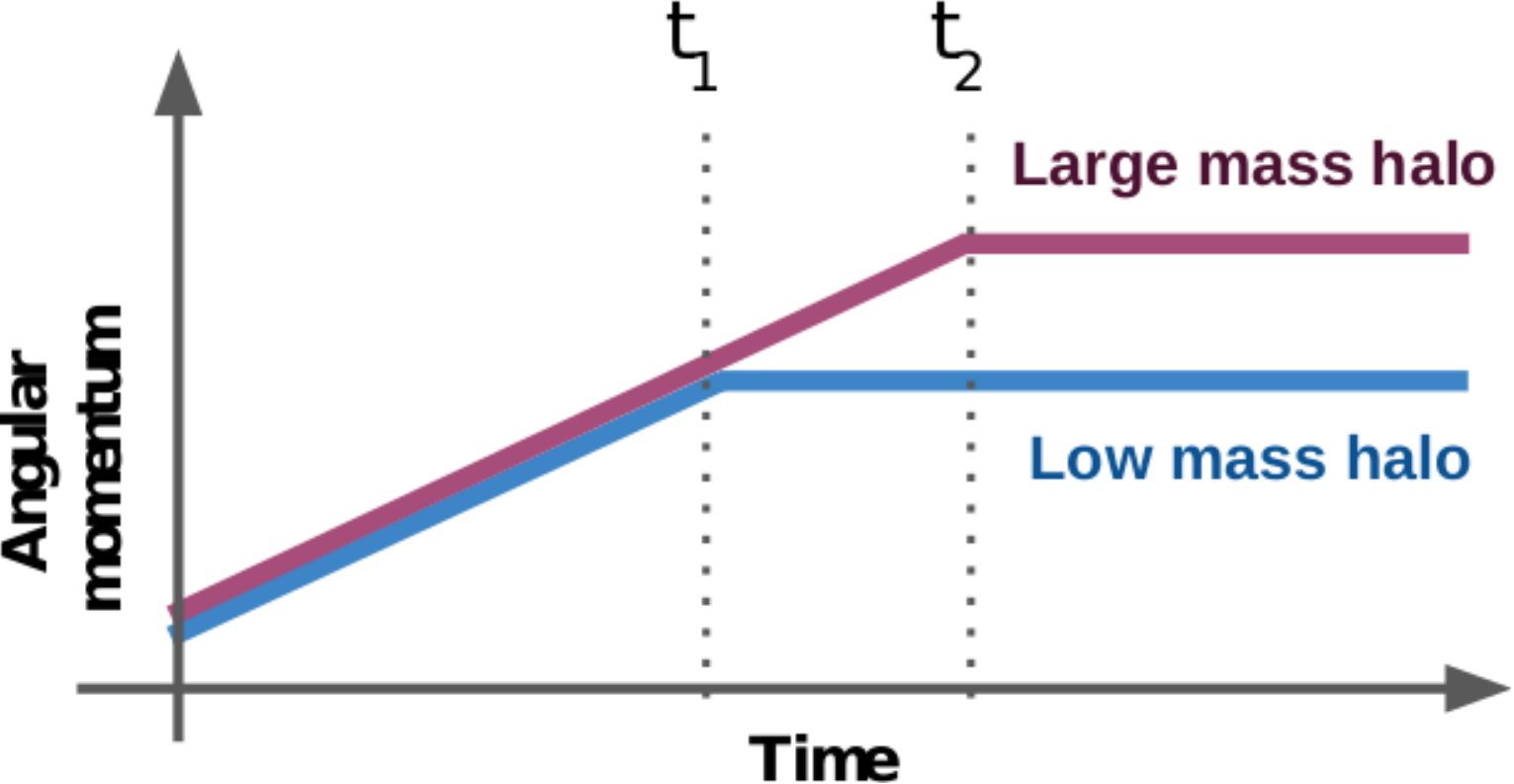}
\caption{Scheme of the expected angular momentum acquisition for two haloes with different masses and comparable initial values in the TTT scenario. During the linear and quasi-linear stages, haloes increase their angular momentum proportionally to the cosmic time $t$. When they reach their turnaround times, $t_1$ and $t_2$, both cease to gain angular momentum by tidal torquing. The large mass halo, which collapses later, reaches a larger final value.}
\label{fig:ta_comparison}
\end{figure}

Since at early times the Universe can be well approximated by the Einstein-de Sitter (Eds) model, the growth rate of the angular momentum $a^2\dot{D}$ can be assumed to be $\propto D^{3/2}$, and therefore behaves approximately as $t$. Thus, the TTT predicts that the angular momentum of DM haloes increases proportionally to $t\propto a^{3/2}$. This growth is expected to stop near the turnaround time, when the neighboring disturbances lose their influence on the protohalo due to its collapse and consequent distancing. From this point on, if there are no other effects taken into account, the angular momentum should remain constant until the present time $t_0$. As we know by the hierarchical theory, large mass haloes collapse later in time, so we should expect their final angular momentum to be larger than for low mass haloes (Figure \ref{fig:ta_comparison}).

\section{Data and halo classification}
\label{metodos}

\subsection{Simulation data}
\label{simulations}
In this work we use a periodic simulation of $1600^3$ DM particles in a computational box of $400\hMpc$ side, with {mass resolution $m_{\rm p}= 1.18219 \times 10^{9}\hMsun$}. The cosmological parameters for a flat low-density universe were taken from the Planck Collaboration results \citep{plankcollaboration2018}, with a matter density $\Omega_{\rm m}=1-\Omega_{\Lambda}=0.315$, Hubble constant $H_{0}=67.4$ km s$^{-1}$ Mpc$^{-1}$, and normalization parameter $\sigma_{8}=0.811$. The identification of particle clumps (namely, DM haloes) has been carried out at the present time by means of a standard friends-of-friends (FOF) algorithm with a percolation length given by $l=0.17$ $\bar{\nu}^{-1/3}$, where $\bar{\nu}$ is the mean number density of DM particles. The simulation has been performed using the second version of the GADGET code developed by \citet{springeletal2005}. We have identified over $6.8\times10^6$ DM haloes, but in order to ensure an accurate measurement of the angular momentum, and for avoiding biases in the determination of certain dynamical properties due to low number of particles (see, for example, \citeauthor{bettetal2007} \citeyear{bettetal2007}), we perform all of our analysis with haloes with at least {$250$} particles, which correspond to a mass {$M\geqslant 3\times 10^{11}\hMsun$}.

For the above described sample of haloes we compute the eigenvalues and eigenvectors of the shape tensor \citep{pazetal2006}{, i.e. the detraced part of the total inertia tensor:}
\begin{equation*}
    \mathbfss{I}_{ij}=\frac{1}{N_h}\sum_{\alpha=1}^{N_h}X_{\alpha i}X_{\alpha j},
\end{equation*}
{where $X_{\alpha i}$ represents the $i$-th component of the displacement vector of the particle $\alpha$ with respect to the centre of mass of the halo, and $N_h$ is the number of particles in the halo. We also compute} the following dynamical properties: angular momentum and spin parameter \citep{bullocketal2001}. In order to study the evolution of the angular momentum, we track the particles of each halo back in time up to $z \sim 80$, computing at every time the properties of the corresponding protohaloes.

\subsection{TTT implementation and halo samples}
\label{ttt_implementation}
There is no unique way to study the TTT predictions in numerical simulations. The turnaround time of a given halo, for instance, can be defined in different ways, and this ambiguity yields considerable uncertainties in the expected final angular momentum. \citet{sugermanetal2000}, for example, use four independent methods to determine the turnaround time: two empirical, computing the time at which the mean radial velocity of the halo particles becomes negative and, alternatively, when the fraction of infalling particles is at least of $50$ per cent; one semi-empirical, measuring the overdensity of the collapsing structure over time until it reaches a certain value given by the spherical collapse model; and the fourth analytical, determining the time at which the linearly extrapolated initial overdensity reaches the value $\delta\sim 1.06$ \citep{peebles1980}. The latter estimate is further adjusted by \citet{porcianietal2002a}, in order to achieve that the average value of the quotient between measurements and predictions approaches unity, thus obtaining that the appropriate time to stop the TTT process is, in fact, approximately half the turnaround time. This method of correction taking into account the proximity between measurements and predictions is also used with other definitions, such as the time at which smoothed protohaloes are predicted by the Z'eldovich approximation to collapse along their major axis. In all cases, due to the major dependence on the stopping point of the TTT process, there is a great uncertainty associated with the determination of the predicted final angular momentum.

We intend to analyze statistically the environmental and dynamical properties of halo samples classified according to their systematic deviations from the model. In order to do so, we should consider that the ambiguity in the definition or computation of certain parameters, such as the turnaround time, could introduce biases in our classification. Consequently, the following implementation is an attempt to identify a common path, as derived from the TTT formalism, in the expected angular momentum evolution of every halo, minimizing its dependency on ambiguous definitions.

\begin{figure}
	\includegraphics[width=\columnwidth]{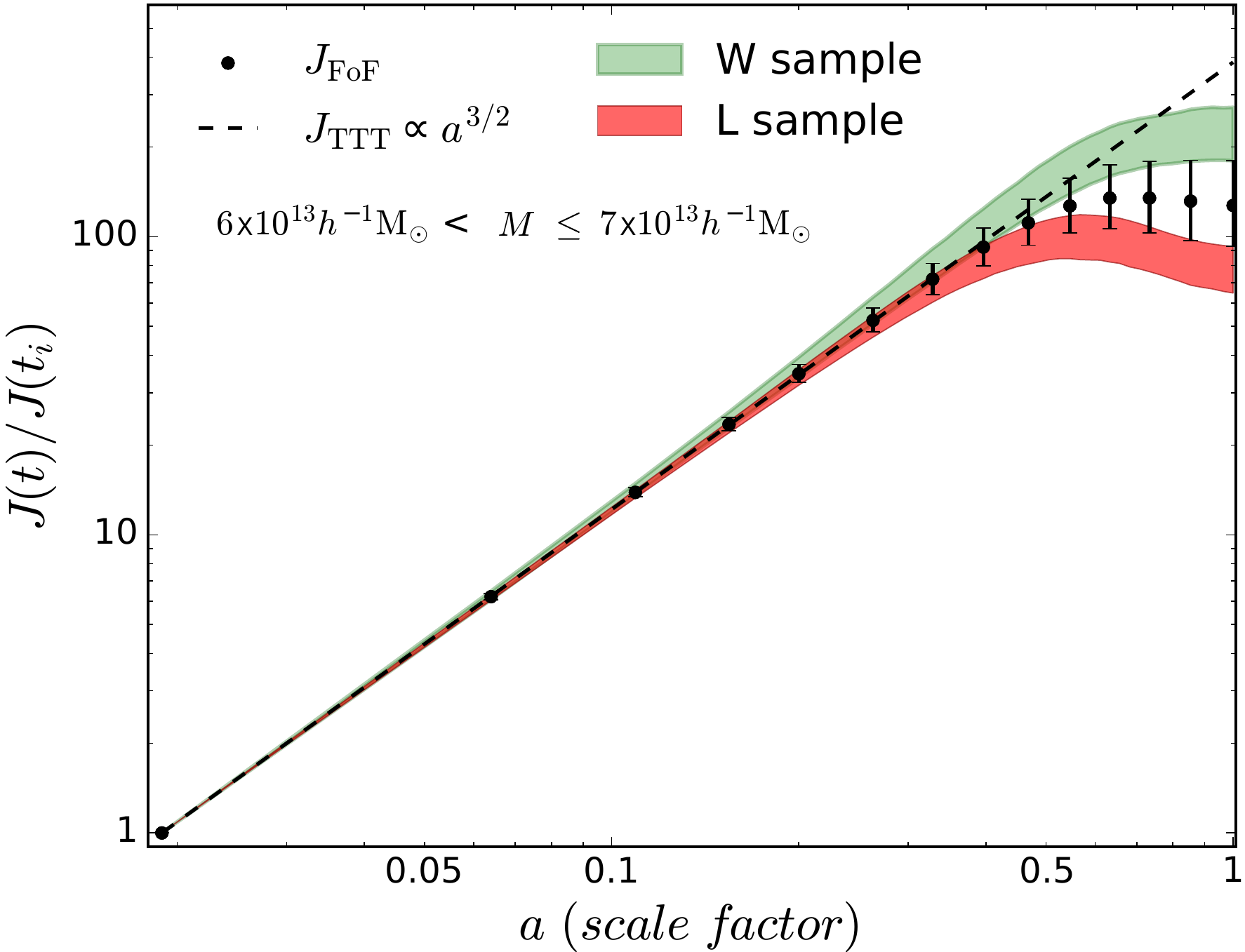}
    \caption{Evolution of the median angular momentum $J_\mathrm{FoF}$ (black circles) for DM haloes in the mass range $[6-7]\times 10^{13}\hMsun$, compared with the predicted evolution $J_\mathrm{TTT}$ (dashed line) in our TTT implementation. The error bars and colored areas show the interquartile range of $J_\mathrm{FoF}$ and the $\Wh$ and $\Lh$ samples at every time, respectively.}
    \label{fig:J_median}
\end{figure}

First, notice that in equation \eqref{ttt_equation} the time-independent factor $\epsilon_{ijk}T_{jl}I_{lk}$ can be assumed to be proportional to the initial angular momentum $J(t_i)$, whose subsequent evolution during the linear stages of structure formation is moduled by the time-dependent factor $a^2(t)\dot{D}(t)$. Since $J(t_i)$ can be directly computed in the initial conditions, we define the {normalized angular momentum} $J_\mathrm{n}(t)=J(t)/J(t_i)$, i.e. the quotient between the {total} angular momentum at each time and its initial value, {regardless the direction of these magnitudes}. Thus, every halo in the simulation has an initial normalized angular momentum $J_\mathrm{n}(t_i)=1$ which is expected to grow during the linear regime as $t\propto a^{3/2}$, until the halo reaches its turnaround point, stops the angular momentum acquisition by tidal torquing and freezes it until the present time.

An statistical analysis of the time evolution of $J_\mathrm{n}$ in halo samples of different masses indicates that its median value follows accurately these predictions, specially during the linear regime. In Figure \ref{fig:J_median} we can see an example of this behavior for DM haloes with masses in the range $[6-7]\times 10^{13}\hMsun$ (black filled circles). Although a tight agreement between the median value and the model can be observed, the scatter (represented by the error bars) increases as we approach the present time, suggesting that non-linear effects begin to play a leading role in the mechanism of angular momentum acquisition. Hence, as noted by \citet{porcianietal2002a}, although the TTT describes accurately the median angular momentum evolution, it is not able to account for the evolution of each individual halo.

In order to identify systematic deviations in this scatter, we use the value of $J_\mathrm{n}(t_0)=J_\mathrm{n,0}$ (i.e. the $J_\mathrm{n}$ value at the present time) as an estimator that allows a rough comparison of the {net angular momentum growth} of haloes. However, as is shown in Figure \ref{fig:ta_comparison}, we expect this quantity to increase with mass, because the more massive a halo is, the later it collapses. To make our analysis independent of this trend, the $J_\mathrm{n,0}$ comparison is performed inside narrow equal-number bins of mass, within which the collapse times are expected to be alike. This allows us to distinguish haloes that, throughout their formation history, have somehow {won} or {lost} angular momentum with respect to what is expected from the TTT, regardless of their initial angular momentum and mass. Consequently, we classify as type $\winner$ ($\Wh$) or $\loser$ ($\Lh$) those haloes that fall, respectively, into the upper or lower terciles of the $J_\mathrm{n,0}$ distribution of their own mass bin. Therefore, by construction, both samples have the same number of haloes (a third from the total population) and very similar mass functions. Additionally, with all the central terciles we define a third sample of $\median$ ($\Mh$) haloes, which in our implementation represent the part of the population whose angular momentum has grown approximately as the TTT predicts. It is important to stress that $\Wh$ and $\Lh$ haloes do not have higher or lower angular momenta, but higher or lower net angular momentum growths. 

\begin{figure}
	\includegraphics[width=\columnwidth]{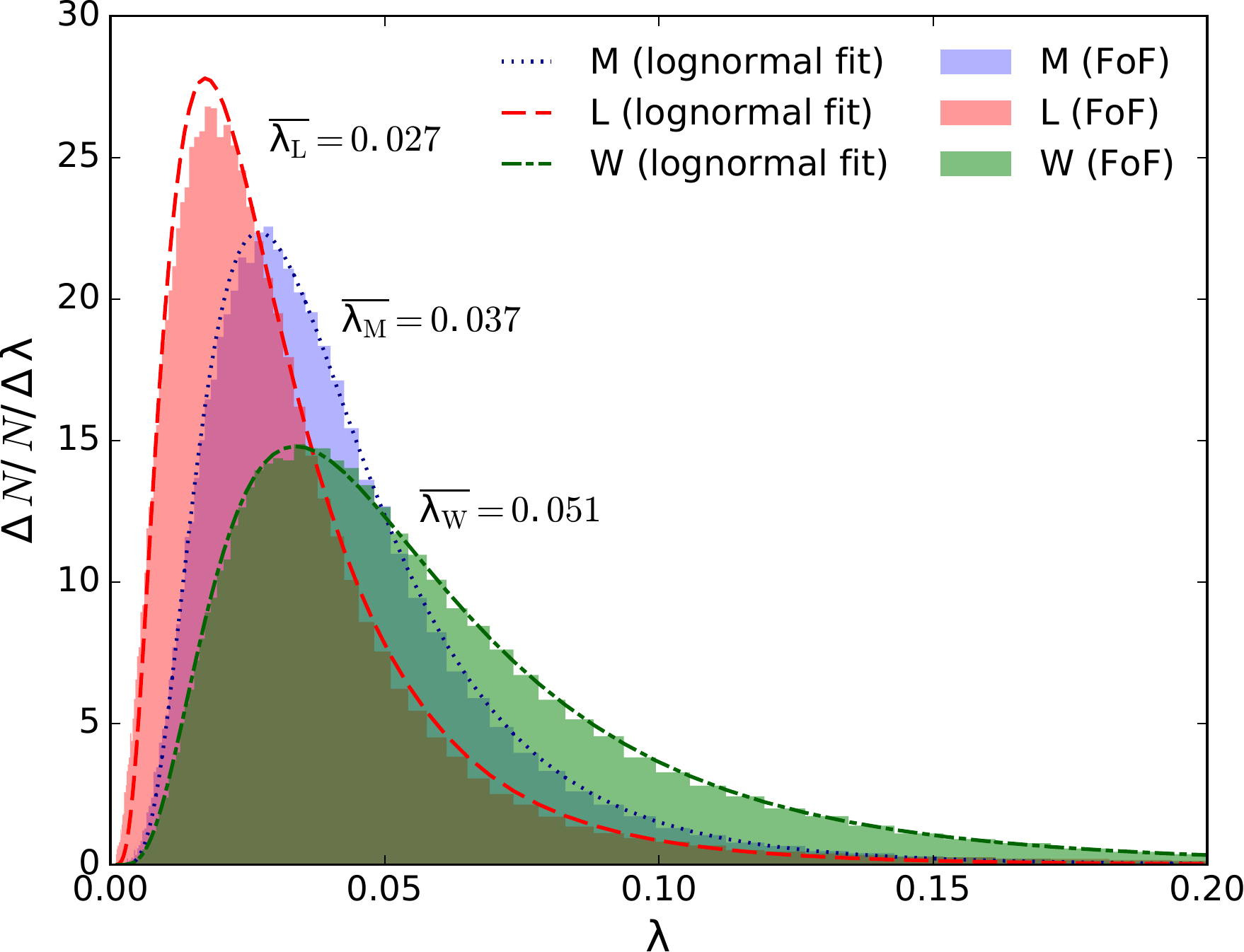}
    \caption{Distribution of the dimensionless spin parameter $\lambda$ at $z=0$. The red, blue and green areas show measures of the $\lambda$ distribution for haloes in the $\Lh$, $\Mh$ and $\Wh$ samples, respectively. The dashed, dotted and dot-dashed curves show the best fitting log-normal distributions for each sample, from which we obtain the peaks $\overline{\lambda_\mathrm{L}}$, $\overline{\lambda_\mathrm{M}}$ and $\overline{\lambda_\mathrm{W}}$.}
    \label{fig:gp_spinparameter}
\end{figure}

To illustrate how DM haloes acquire angular momentum over time according to this classification, Figure \ref{fig:J_median} also includes the evolution of the $\Wh$ and $\Lh$ samples as green and red shaded areas, respectively. To that end, we have computed the interquartile difference of $J_\mathrm{n}$ within each sample at every time. By construction, these are the external terciles of the $J_\mathrm{n}$ present-day distribution, but as we look back in time they tend progressively to the median behavior. During the linear regime, when $J_\mathrm{n}$ grows proportionally to $a^{3/2}$, the samples $\Wh$ and $\Lh$ are tightly attached to the model predictions although they show a slight bias towards positive or negative deviations, respectively. This suggests that the effects that produce large differences in the final stages of structure formation could have been present before the haloes began to collapse.

\section{Analysis}
\label{results}
\subsection{Dynamical properties of the halo samples}

\begin{figure}
	\includegraphics[width=\columnwidth]{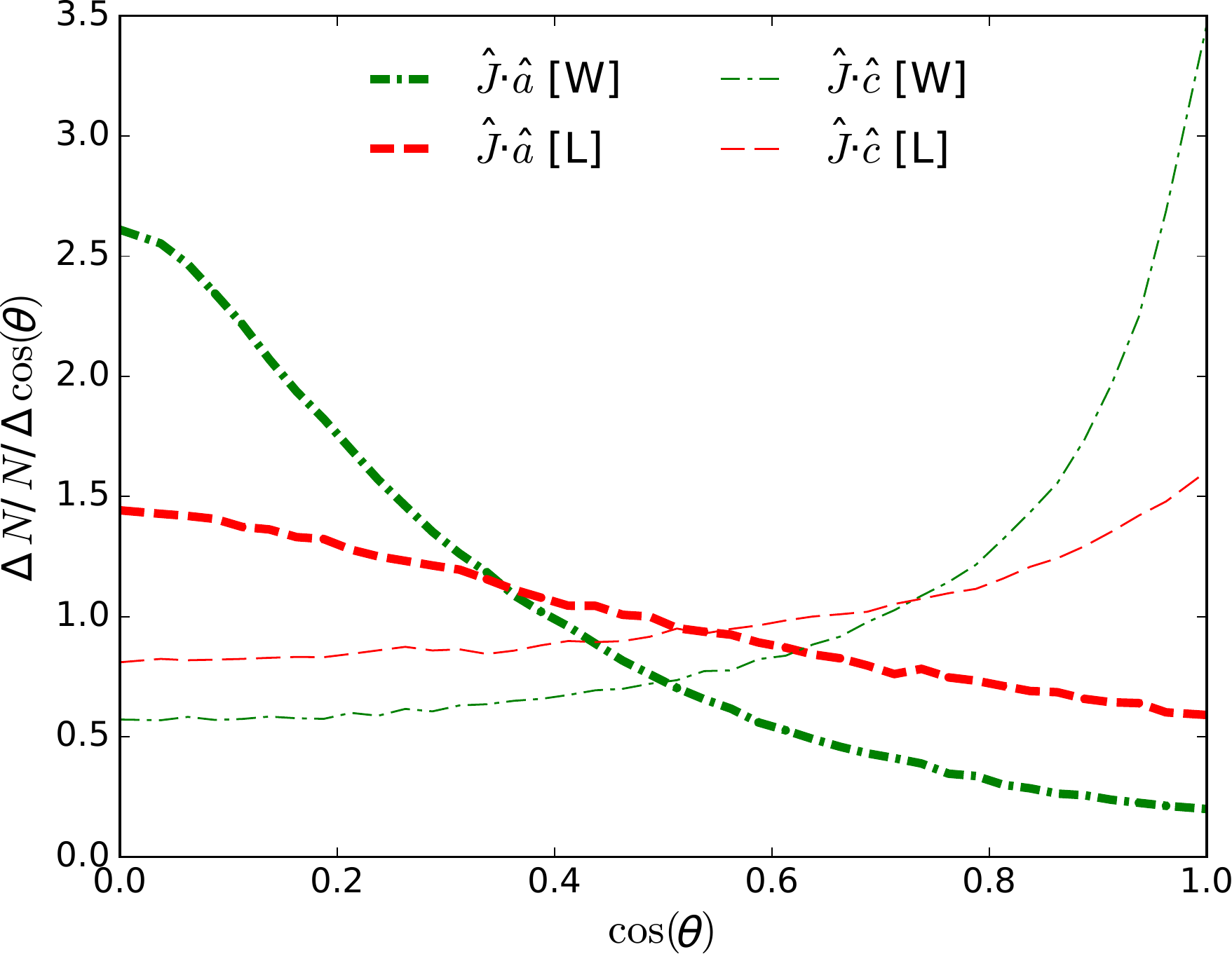}
    \caption{Distribution of the cosine of the angle $\theta$ between the halo angular momentum and the principal axes of the shape tensor at $z=0$. The perpendicularity increases towards the left side of the panel, while the parallelism augments towards the right. The thick (thin) curves show the alignment between the major (minor) axis $\hat{a}$ ($\hat{c}$) and the angular momentum direction $\hat{J}$. The red dashed and green dot-dashed lines correspond, respectively, to the $\Lh$ and $\Wh$ samples.}
    \label{fig:histo_Jxejes}
\end{figure}

As it was mentioned above, our classification according to the angular momentum growth is motivated in the analysis of systematics on the TTT deviations over the halo population. At a first glance, this criterion may seem rather arbitrary. However, the following analysis of the internal properties of the samples indicates that $\Wh$ and $\Lh$ haloes are, in fact, dynamically different structures. In a first approach, we can see in Figure \ref{fig:gp_spinparameter} the present-day distribution of the dimensionless spin parameter $\lambda$ for haloes in the three samples. It is clear that each distribution follow a log-normal trend, and that $\Wh$ and $\Lh$ haloes are shifted from the distribution of the $\Mh$ sample, which is consistent with the well known peak at $\lambda_{0}\sim 0.04$. While $\Lh$ haloes have typical spin parameters $30$ per cent lower than $\lambda_\mathrm{0}$, the $\Wh$ sample peaks at a value $27$ per cent above it, and about the double as the $\Lh$ sample. This indicates that there is a strong correlation between the coherent rotation inside DM haloes and our classification according to the deviations from the TTT, which could be roughly stated as the greater the angular momentum growth, the higher the rotation support of the halo. {Consistently with the results found in previous works, this trend depends rather weakly on the halo masses.}

\begin{figure*}
	\includegraphics[width=2\columnwidth]{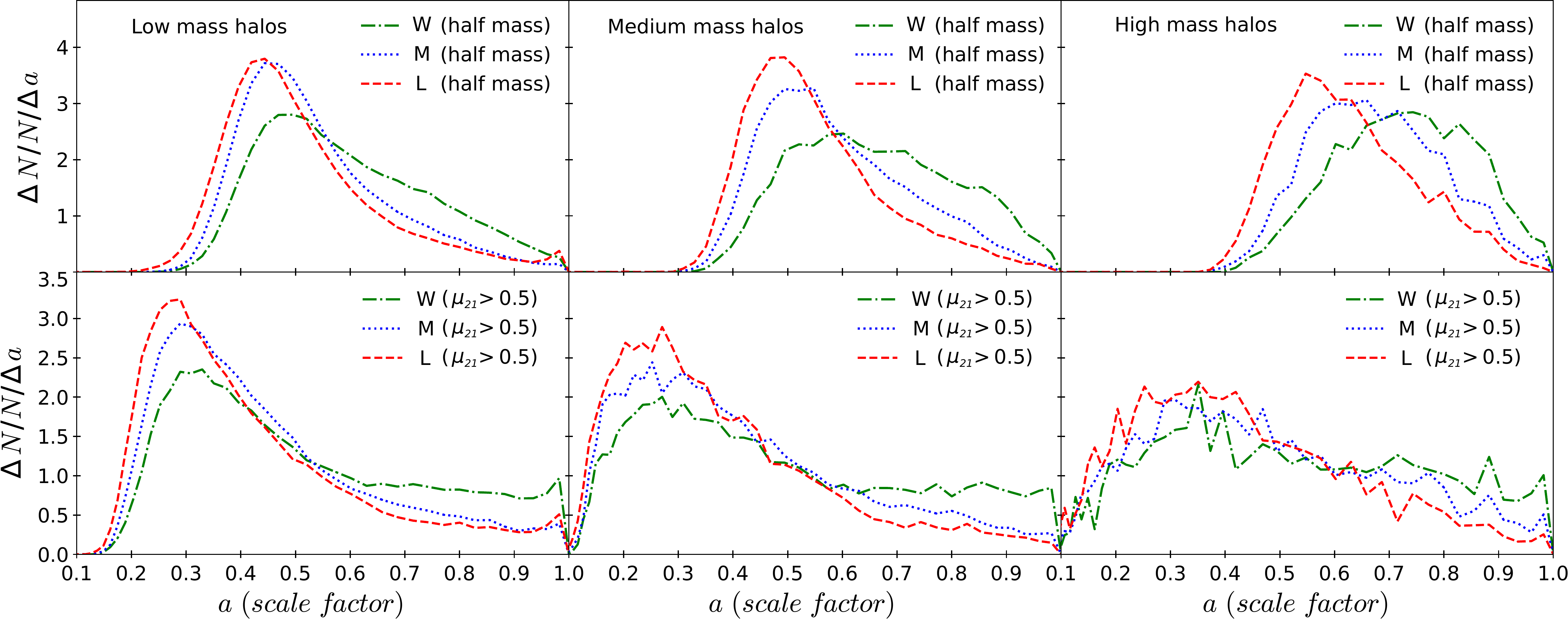}
    \caption{{Distribution of the scale factors at which the mass of the main branch of each DM halo reaches half of its present day mass (upper row) and at the last time at which the ratio between the second most massive progenitor and the main progenitor is equal or greater than $0.5$  (lower row). The three columns show increasing mass ranges, from left to right, and the samples of $\Wh$, $\Mh$ and $\Lh$ haloes are represented, respectively, by green dot-dashed, blue dotted and red dashed lines.}}
    \label{fig:acrecion_M123}
\end{figure*}

Another analysis that supports the relevance of our classification can be found in the internal alignment of haloes. In Figure \ref{fig:histo_Jxejes} we show the present-day distribution of $\mathrm{cos}(\theta)$, where $\theta$ represents the angle between the angular momentum and the major (thick curves) and minor (thin curves) axes of the shape tensor, for haloes in the $\Wh$ and $\Lh$ samples. There is a clear tendency for $\Wh$ haloes to have their angular momentum direction correlated with their shape. In effect, we see that, for most of $\Wh$ haloes, $\hat{J}$ is strongly aligned with the minor axis $\hat{c}$, and perpendicularly oriented with respect to the major axis $\hat{a}$. This trend is consistent with the positive shift observed in the $\lambda$ distributions of the $\Wh$ sample, since haloes with higher rotational support are less likely to collapse in the plane perpendicular to their angular momentum. In contrast, $\Lh$ haloes show a rather weak internal alignment, with much more uniformly distributed values of $\mathrm{cos}(\theta)$. 

\subsection{Accretion history}

{In order to understand the possible origin of the differences in our samples, we build merger trees and analyze how our classification can be related to the haloes assembly history. To this end, in this section we refer to the present day halo population as root haloes, and divide them in three groups according to their mass: $[3 - 7]\times10^{11}\hMsun$ (low mass), $[3 - 7]\times10^{12}\hMsun$ (medium mass) and $[3\times10^{13} - 2\times10^{15}]\hMsun$ (high mass). 
The merger trees are constructed by performing FOF identifications across the entire simulation, with the same percolation length and mean number density of DM particles that we used in the first run.
However, given that root haloes are classified according to their net angular momentum growth, and that this is computed with no gain or loss of the present day mass, the FOF identifications are carried out considering and connecting together only particles that have been associated to present day haloes. Then, at each snapshot, we define the set of structures that share at least $70$ per cent of their particles with a given root halo as its progenitors. For consistency, the mass of each progenitor is determined considering only the fraction of particles that end up in the correspondent root halo\footnote{Due to the selection criteria, the results show no differences if we compute the mass of the progenitors using always $100$ per cent of their particles.}. The most massive progenitor at each time is then defined as the main progenitor, and the evolution of these haloes form the main branch of each merger tree.}

{We study two types of events in the haloes assembly history. On the one hand, we determine the scale factor at which the main branch reaches half the mass of the root halo. This estimator indicates the moment from which most of the present day mass of haloes is concentrated in a single structure, and therefore can be related with their collapse time. On the other hand, we analyze the evolution of the ratio $\mu_{21}$ between the mass $M_2$ of the second most massive progenitor and the mass $M_1$ of the main progenitor, and determine the scale factor at the last time that $\mu_{21}\geqslant 0.5$. This gives us a complementary measure of the mass distribution in the evolution of haloes. Firstly, if the majority of the mass reaches recent times distributed in two large structures, it is to expect that the tidal interaction with the surrounding structure is more effective. Secondly, the earlier $\mu_{21}$ drops below $0.5$, the lower the probability that the main progenitor has suffered major mergers that significantly affect the final angular momentum of the halo.}

\begin{figure*}
	\includegraphics[width=2\columnwidth]{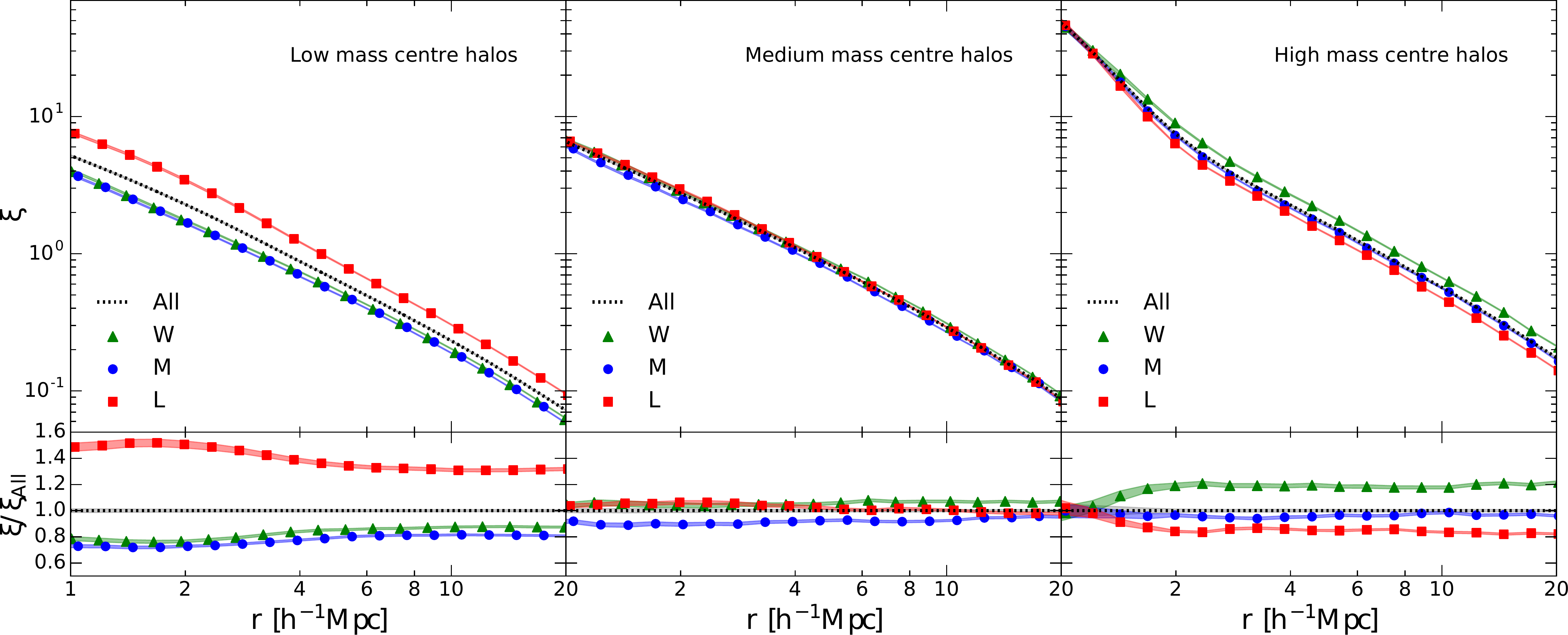}
    \caption{Measures of isotropic halo-halo correlation functions. The upper panels show the correlation between centre haloes of low, medium and high mass, from left to right, and neighbors of all masses. These values are presented as a function of the distance from the centre haloes. The green triangles, blue circles and red squares correspond, respectively, to the correlation using $\Wh$, $\Mh$ and $\Lh$ haloes as centres, while the black dotted curve show $\xi_\mathrm{All}$, i.e. the results using all centres in each mass range, regardless the sample they belong. In shaded areas of the corresponding colors we present the error estimations, which in most cases appear to be thin lines. The bottom panels show the quotients between the results shown in the upper panels and $\xi_\mathrm{All}$.}
    \label{fig:iso_M123}
\end{figure*}

{The probability density of these two measurements are presented in Figure }\ref{fig:acrecion_M123}{: the scale factors at half mass (upper row) and at the last time that $\mu_{21}\geqslant 0.5$ (lower row), for haloes in the three mass ranges (increasing from left to right) and separated according to our classification (green dot-dashed, blue dotted and red dashed curves for $\Wh$, $\Mh$ and $\Lh$ haloes, respectively). 
The scale factor at half mass shows a significant dependence on the net angular momentum growth. For instance, the median value for high mass $\Lh$ haloes can be found at $a=0.60$, whereas the median of the $\Wh$ sample is $19$ per cent higher, at $a=0.71$. As can be seen, the shape of the distribution of $\Wh$ haloes continuously changes with mass, showing a persistent tendency to assemble half of their mass more recently.
In contrast, both the shape and the peak of the distributions for $\Lh$ haloes show a weaker dependence on mass. In addition, it can bee seen that $\Lh$ haloes show earlier formation times when compared to $\Wh$ haloes.
Regarding the distributions of the last time that $\mu_{21}\geqslant 0.5$, our results show that, for all masses and samples, there is a peak between $a\sim 0.3$ and $a\sim 0.4$. However, as we approach recent times, the $\Wh$ sample shows a persistent fraction of haloes whose mass is still distributed among two comparable structures, whereas this fraction is significantly smaller for the $\Lh$ and $\Mh$ samples.}

{These results suggest that the assembly history of DM haloes is significantly correlated to their angular momentum acquisition. 
While $\Lh$ haloes accrete most of their mass at early times, a significant part of the $\Wh$ sample do so more recently.
In addition, there is a large number of $\Wh$ haloes whose second most massive progenitor still represents, at recent times, a considerable fraction of the mass of their main progenitor.
Therefore, near the present time, $\Wh$ haloes are more fragmented and their mass more dispersed. 
This suggest that their inertia tensor may not have completely decoupled from the tidal tensor of the surrounding matter distribution, even when they have entered into the non-linear regime, and could be one of the reasons for their higher angular momentum growths.} 

\subsection{The spatial halo-halo correlation function}

Considering that the classification according to deviations from the TTT has resulted in halo samples with significantly different dynamical properties {and assembly history}, it is worth asking about the role that the surrounding LSS plays in the development of these deviations. To that end, in this section we present measurements of the spatial halo-halo correlation function, using pairs of DM haloes in the $\Lambda$CDM simulation introduced in Section \ref{metodos}. In order to achieve a better understanding of how the clustering affects the angular momentum growth, we focus on the differences arising from taking into account the halo masses and belonging samples.

The spatial halo-halo correlation function $\xi(r)$ measures the excess probability $dP$ of finding a halo in a volume element $dV$, at a distance $r$ away from a given centre halo, with respect to a random distribution. This can be expressed as
\begin{equation*}
dP=\bar{n}[1+\xi(r)]dV,
\end{equation*}
where $\bar{n}$ is the mean number density of DM haloes in the simulation. To compute $\xi$ we count, for a given halo sample (centre haloes), the number of neighboring objects (tracers) found at different distance bins. The total number of neighbors per interval, $DD$, is then normalized by the number of pairs expected in a homogeneous distribution, $DR$. Finally, for each distance $r$, the excess with respect to the unit of the stacked count is our estimator of the correlation function $\xi(r)=DD(r)/DR(r)-1$. The error estimations in the determination of the correlation functions are obtained through a jackknife algorithm with a total of $100$ subsamples\footnote{We tested from 2 to 300 subsamples and found that our error estimations stabilize around 100.}. In the case of quotients between correlation functions, we propagate these uncertainties through a partial derivatives analysis.

In order to statistically analyze how the LSS affects the angular momentum growth, our computations are performed using as centre haloes the samples defined in Section \ref{metodos}, $\Wh$, $\Lh$ or $\Mh$. Additionally, we compute correlation functions using centre haloes regardless their belonging sample, i.e. haloes of the {general population}. We also consider the mass in our study. 
Although the counting of neighbors is always done over the full range of halo masses, {centre haloes are classified, as defined in the previous section, in the three ranges labeled as low, medium and high mass.}
In Figure \ref{fig:iso_M123} we present measures of the halo-halo correlation function for these mass ranges. Each panel has four curves: the triangles, circles and squares correspond, respectively, to the correlation function taking centre haloes from the samples $\Wh$, $\Mh$ and $\Lh$, and the black dotted curve shows $\xi_\mathrm{All}$, the correlation function using all centres of the given mass range, regardless the sample they belong. The error estimations are represented by the shaded areas, which in most cases appear to be thin lines. The bottom panels show, in the same order and symbol key, the quotients between each correlation function and the corresponding $\xi_\mathrm{All}$.

At a first glance in the upper panels, it can be noticed that the correlation function of the general population tends to increase, as it is expected, with the mass of the centre haloes. 
This behavior is more noticeable when we consider $\Wh$ haloes. For this sample, the parameter $r_0$ of the best fitting power law $\xi(r)=(\frac{r}{r_0})^{\gamma}$ changes from $\sim 3.3\hMpc$ at low masses to $\sim 7.7\hMpc$ at high masses, whether the correlation function for haloes in the $\Lh$ sample exhibit a rather weak dependence on mass, with $r_0$ barely changing from $\sim 4.6\hMpc$ to $\sim 6.1\hMpc$. 

A noteworthy feature in the high mass range (upper right-hand panel) is the abrupt change in the slope that occurs near $2\hMpc${, which looks very similar to the} characteristic jump between the one- and two-halo terms, according to the halo model \citep{coorayysheth2002}. {Given that we have computed our correlation functions using halo-halo pairs, the one-halo term reveals the matter distribution in the immediate vicinity of high mass haloes, whereas the two-halo term traces the surrounding LSS. Hence, the overlapping of the $\Wh$, $\Mh$ and $\Lh$ correlation functions below $2\hMpc$ suggests that the most likely responsible for the systematic deviations from the TTT should not be sought in the immediate vicinity of high mass haloes, but in the large scales instead. This being said, we will focus our analysis on the two-halo term. As it is expected, the $\Mh$ sample shows a typical surrounding structure very similar to that of the general population. More interestingly, the $\Wh$ and $\Lh$ samples exhibit opposite shifts from $\xi_\mathrm{All}$}. Massive haloes with high angular momentum growth are embedded in highly clustered environments, while haloes with similar masses but low angular momentum growth inhabit somehow less clustered regions. Although high mass haloes typically form the densest structures of the cosmic web (namely, the nodes of the LSS), this trend indicates that there is a remarkable bias on how $\Wh$ and $\Lh$ haloes occupy these regions. This, in turn, suggests that a post-TTT process may be taking place during the final stages of structure formation, enhancing the angular momentum acquisition of massive haloes with a certain excess of neighbors. 

\begin{figure}
	\includegraphics[width=\columnwidth]{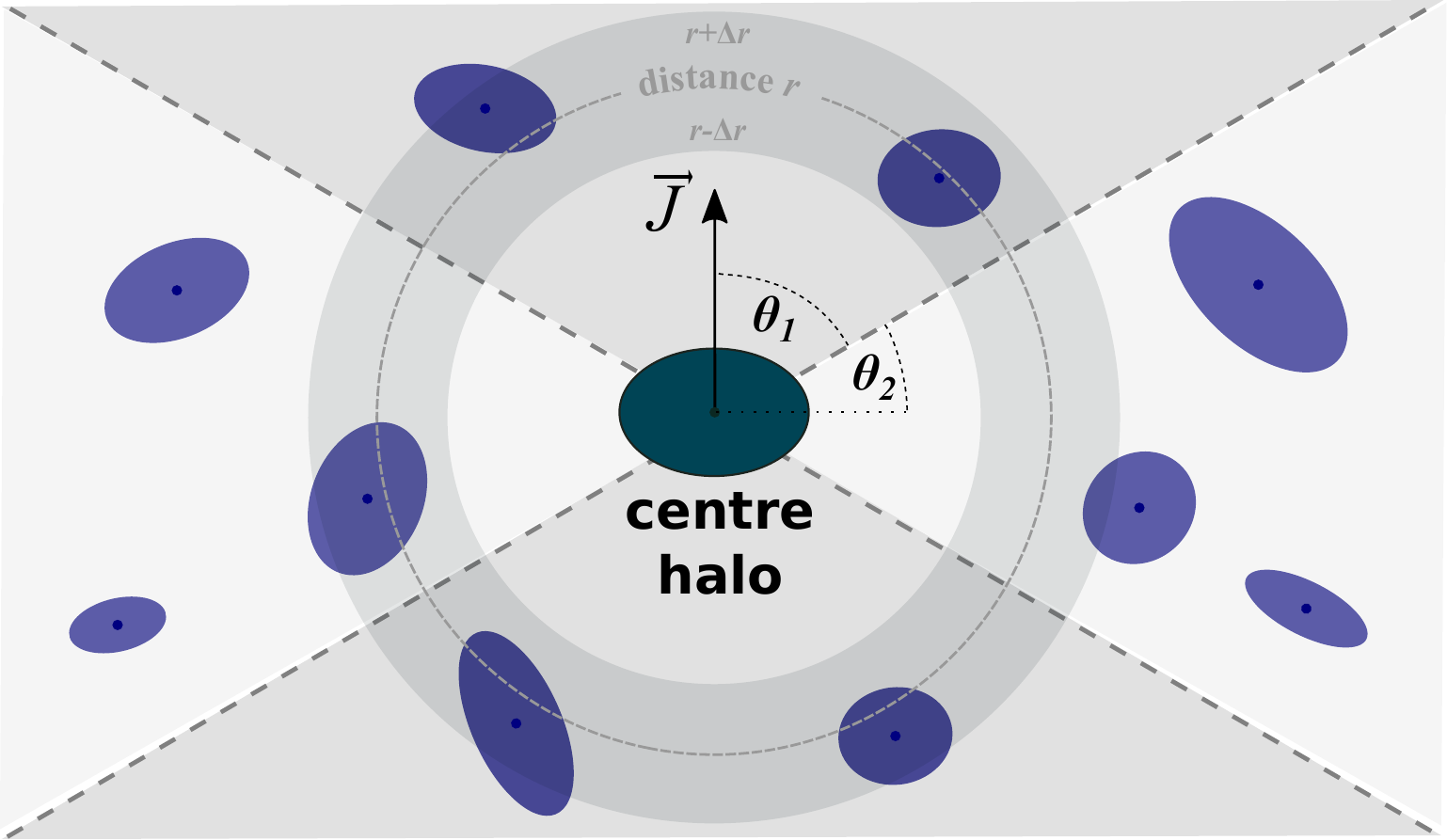}
    \caption{Two dimensional sketch of the neighbors in the vicinity of a centre halo with angular momentum $\mathbfit{J}$. The dark and light shaded areas represent, respectively, the volumes parallel and perpendicular to $\mathbfit{J}$, defined by the threshold angles $\theta_1$ and $\theta_2$. The gray circle represent the spherical shell used to compute $\xi$ at distance $r$, whereas its intersection with the shaded areas represent the regions where the pairs contribute to $\xi_{\parallel}$ and $\xi_{\perp}$.}
    \label{fig:esquema_fca}
\end{figure}

As we look toward lower masses, we can see that the quotients with $\xi_\mathrm{All}$ (lower panels) show an inversion of the relative differences of the $\Wh$ and $\Lh$ correlation functions. At high masses, $\Wh$ haloes show an excess of correlation of about $20$ per cent, while in the lower masses they show a clear lack of correlation, from $10$ per cent to $20$ per cent below $\xi_\mathrm{All}$. Conversely, high mass $\Lh$ haloes show a signal about $20$ per cent below the general population, but at low masses their surrounding structure is remarkably more clustered than the rest of the population of the same mass, with an excess of $\sim 40$ per cent above $\xi_\mathrm{All}$, and about the double as the $\Wh$ sample.

The medium mass range shows an almost complete overlap between the correlation functions of the different samples and $\xi_\mathrm{All}$, except for a slight lack of correlation around $\Mh$ haloes as we get closer to  smaller scales. This can be interpreted as a transition regime between what we observe in the higher and lower mass ranges: high mass $\Wh$ haloes typically inhabit over-clustered regions, whereas low mass haloes of the same sample are systematically found in environments less densely populated than usual. The inverse trend occurs to the $\Lh$ sample.

\subsection{Anisotropies in the correlation function}
\label{anisotropic_xi}

\begin{figure*}
	\includegraphics[width=2\columnwidth]{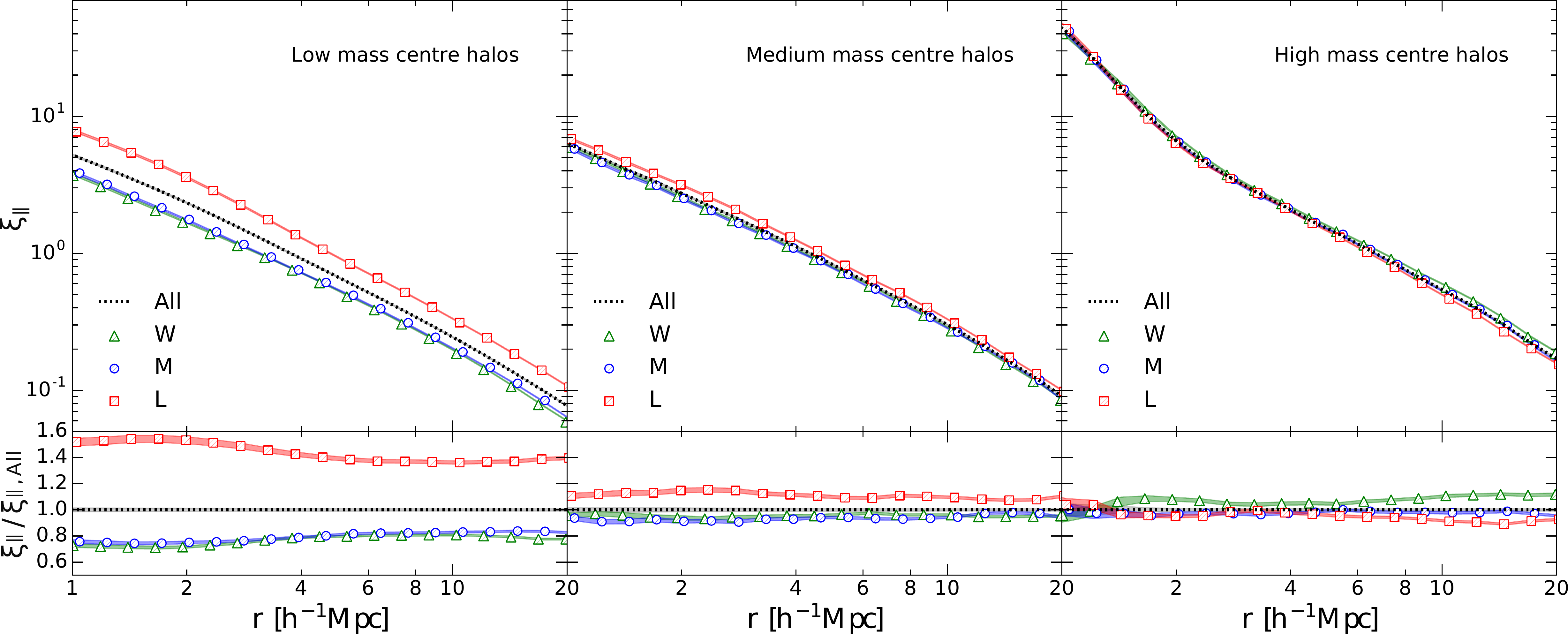}

    \vspace{0.3cm}
    \includegraphics[width=2\columnwidth]{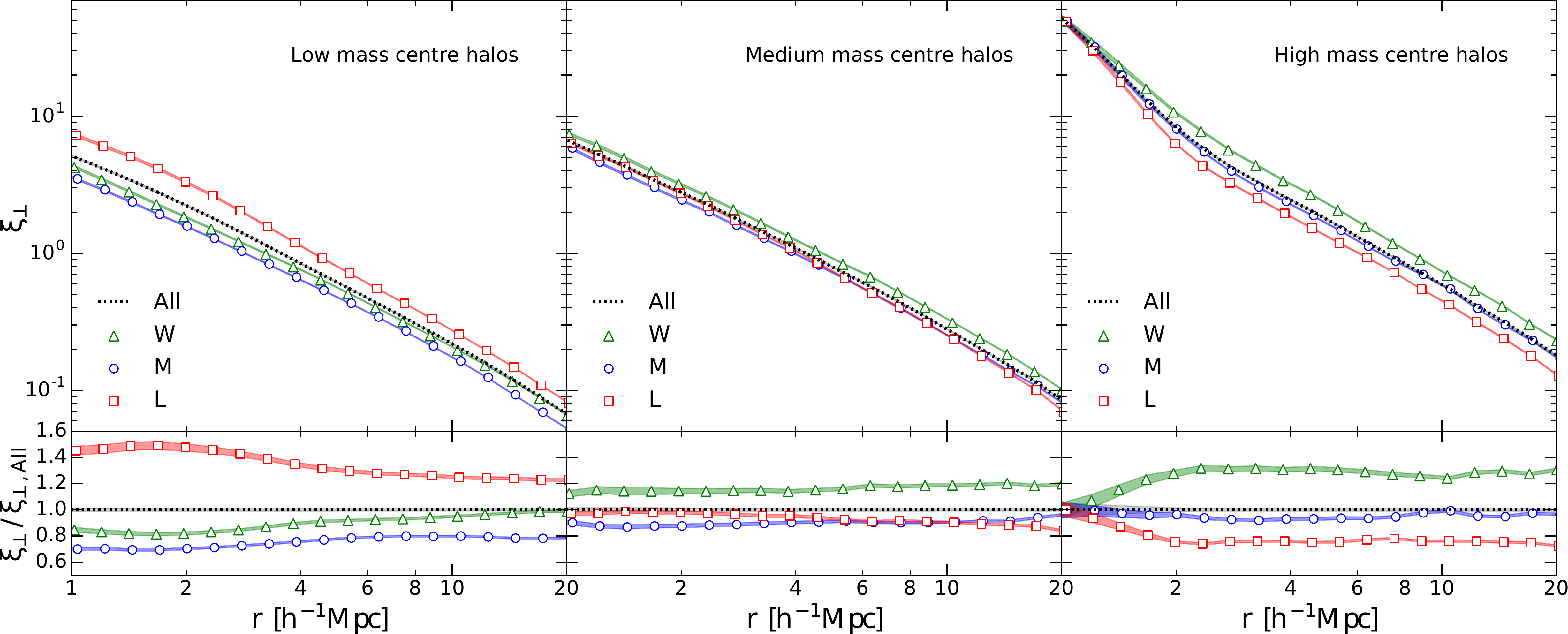}
    \caption{Panels corresponding to the anisotropic correlation functions, analogous to the set shown in Figure \ref{fig:iso_M123}. In the upper panels (shaded symbols), the neighbors count is performed only in the direction parallel to the angular momentum of the centre haloes, while in the lower panels (open symbols) only neighbors in the perpendicular direction are taken into account.}
    \label{fig:ani_M123}
\end{figure*}

In order to understand how the direction of the surrounding structure affects the angular momentum acquisition, in addition to the {isotropic} correlation functions presented in the previous section, we compute {anisotropic} correlation functions. Following \citet{pazetal2008}, instead of counting neighbors around the centre haloes in all directions, we take into account the orientation of the {centre-neighbor} pairs with respect to the angular momentum of the centre haloes. Hence, the anisotropic correlation functions are computed considering two cases:
\begin{itemize}
\item when the pairs subtend an angle from the direction of the  angular momentum of the centre halo less than a given threshold $\theta_1$ ({parallel case}, dark shaded region), 
\item when the pairs are found at lower inclinations than a limit angle $\theta_2$ from the perpendicular plane to the angular momentum ({perpendicular case}, light shaded region).
\end{itemize}
We choose the threshold angles $\theta_1$ and $\theta_2$ so that the volumes in each case are the same. This is achieved by setting $\mathrm{sin}(\theta_2) = 1-\mathrm{cos}(\theta_1) = \chi$, and choosing a value for the threshold parameter $\chi$. Selecting $\chi \leq 0.5$ implies angles $\theta_1 \leq 60\degr$ and $\theta_2 \leq 30\degr$. Throughout this subsection we use $\chi = 0.5$, thus $\theta_1=90\degr-\theta_2=60\degr$.

In Figure \ref{fig:esquema_fca} we present a two dimensional sketch of the neighborhood around a centre halo with angular momentum $\mathbfit{J}$. Every neighbor whose centre of mass is at a distance from the centre halo within the range $[r-\Delta r,r+\Delta r]$ contributes to the isotropic correlation function $\xi(r)$, while the dark and light shaded areas represent, respectively, the regions {parallel} and {perpendicular} to $\mathbfit{J}$, defined by the threshold angles $\theta_1$ and $\theta_2$. Hence, the same procedure as in the isotropic case is followed to estimate the anisotropic correlation functions $\xi_{\parallel}(r)$ and $\xi_{\perp}(r)$, but only haloes in the parallel and perpendicular regions, respectively, are taken into account, with its corresponding normalization.

We present results of these measurements in Figure \ref{fig:ani_M123}. In order to analyze both cases separately, the parallel and the perpendicular, we present two sets of panels, each of which is analogous to the set of panels in Figure \ref{fig:iso_M123}. While the upper set correspond to the parallel case (hatched symbols), the lower set show the results corresponding to the perpendicular case (open symbols). Once again, green triangles, blue circles and red squares represent, respectively, the correlation functions using centre haloes from the $\Wh$, $\Mh$ and $\Lh$ samples. The dotted curves show, for each mass range, $\xi_\mathrm{\parallel,All}$ and $\xi_\mathrm{\perp,All}$, i.e. the anisotropic correlation functions using centre haloes from the total sample. 

At low masses (left-hand panels) there is a clear excess of correlation of neighbors around $\Lh$ haloes, both in the directions perpendicular and parallel to their angular momentum. This excess with respect to the general population tends to increase as the distances become smaller. In the parallel case, the excess goes from almost a $50$ per cent effect at small separations to about $40$ per cent at large separations, while in the perpendicular case this trend is more pronounced, from $\sim 50$ per cent to $\sim 20$ per cent. These results suggest that the effect of highly clustered regions is to inhibit the angular momentum acquisition of low mass haloes, perhaps due to the lack of coherence between the torque exerted by their initial surrounding tidal field and the highly non-linear environments in which they inhabit at the present time. Haloes of the $\Wh$ and $\Mh$ samples show, on the contrary, a noticeable lack of correlation with respect to the general population. Nevertheless, whereas in the case of $\Mh$ haloes this trend does not depend on the direction, $\Wh$ haloes show a higher correlation with neighbors near the plane perpendicular to their angular momentum than along their axis of rotation. Moreover, at distances $\ga 10\hMpc$, the perpendicular correlation of $\Wh$ haloes becomes indistinguishable from $\xi_\mathrm{\perp,All}$, while the parallel correlation remains $20$ per cent below. A possible interpretation of these results, following the reasoning used in the case of $\Lh$ haloes, is that the lower level of correlation around low mass $\Wh$ and $\Mh$ haloes could be due to their surrounding tidal fields not having suffered large variations in the passage from linear stages to non-linear stages. However, the higher clustering around $\Wh$ haloes in the direction perpendicular to the angular momentum suggests that, even in these less dense regions, any excess of neighbors near the plane perpendicular to the axis of rotation acts as an angular momentum enhancer.

In the high mass range, on the right-hand panels of Figure \ref{fig:ani_M123}, the differences between the parallel and the perpendicular cases are noticeable. The parallel correlation functions corresponding to haloes of the $\Wh$ and $\Lh$ samples steadily overlap with $\xi_\mathrm{\parallel,All}$ along the entire distance range. This suggests that there is no correlation between the deviations from the TTT of high mass haloes and the clustering in the direction of their angular momentum. The perpendicular case, on the other hand, shows strongly correlated neighbors around $\Wh$ haloes, and a noticeable lack of structure near haloes of the $\Lh$ sample, whereas the correlation of $\Mh$ haloes overlaps with $\xi_\mathrm{\perp,All}$. This is consistent with the hypothesis of a secondary tidal torque acting on haloes after their turnaround, given that any structure enhancing or sustaining the angular momentum acquisition has to lie preferentially in the direction perpendicular to $\mathbfit{J}$. {The overlapping of the curves below $2\hMpc$ observed in the isotropic case (right-hand panel of Figure }\ref{fig:iso_M123}{) is also present in the perpendicular correlations, which supports the idea that the matter distribution that immediately surrounds high mass haloes does not play a significant role in the development of their systematic deviations from the TTT.}

As we saw in the isotropic analysis, the anisotropic correlation functions for haloes in the medium mass range show a sort of transition regime, both in the parallel and perpendicular case, between the results for higher and lower masses. In effect, the central panels in Figure \ref{fig:ani_M123} show almost not difference between the correlation functions of the three samples, except for a slight preponderance of parallel structure near $\Lh$ haloes, and a modest excess of perpendicular structure surrounding $\Wh$ haloes. A noticeable feature is, however, the consistency of the relation between the correlation functions of $\Wh$ and $\Mh$ haloes along the entire mass range: in the direction of the angular momentum the curves always overlap, whereas in the perpendicular direction $\Wh$ haloes always show a difference of correlation $\ga 20$ per cent. 

\begin{figure}
	\includegraphics[width=\columnwidth]{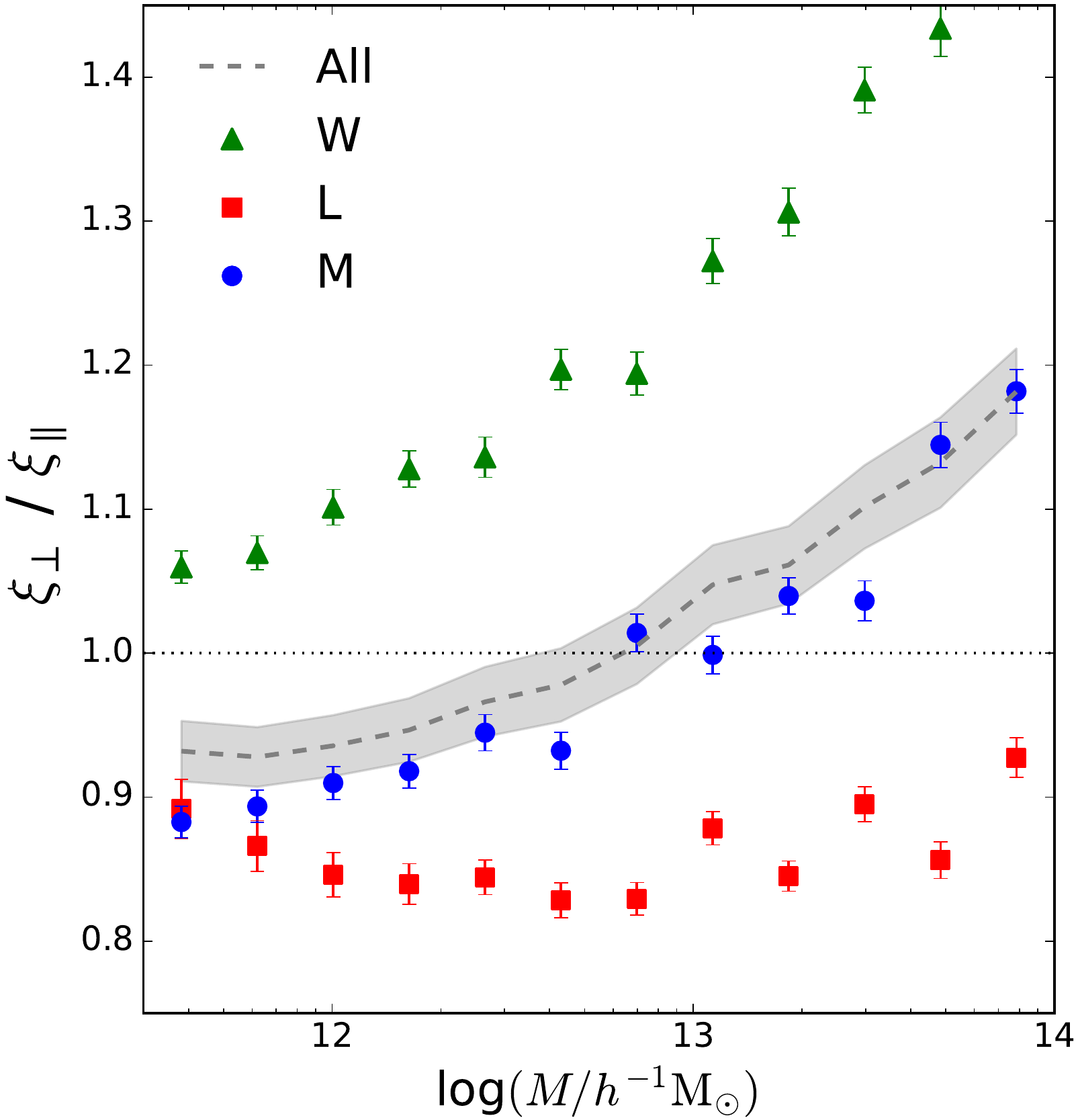}
    \caption{Quotients between the anisotropic correlations $\xi_{\perp}/\xi_{\parallel}$, computed with a single distance range covering the two-halo term, as a function of the centre halo masses. The green triangles, blue circles and red squares show the values corresponding to the $\Wh$, $\Mh$ and $\Lh$ samples, respectively, while the dashed line and gray shaded area show the resulting quotients and their $3\sigma$ dispersion when using centres regardless the sample they belong. Error bars are computed propagating uncertainties through a partial derivative analysis.}
    \label{fig:gp_bines}
\end{figure}

In order to diminish the effect of covariance between the different correlation scales, we perform a single measurement of each anisotropic correlation using all the corresponding halo-halo pairs separated by distances within the two-halo range. In Figure \ref{fig:gp_bines}, we present these measurements through mass dependent quotients between the perpendicular and the parallel cases. These allow to quantify the anisotropy of structure surrounding haloes of a given mass over the two-halo regime. Green triangles show the ratios corresponding to the $\Wh$ sample, blue circles correspond to the $\Mh$ sample and red squares correspond to the $\Lh$ sample. The dashed line and gray shaded area show the quotients and their $3\sigma$ dispersion for the general population. 

As can be seen, there is a clear tendency for the general population to show more structure in the direction perpendicular to the angular momentum as the mass increases. Notice that haloes of the $\Mh$ sample appear to follow the same trend, although with some scatter. This suggests the existence of a ``spin flip'' or ``transition mass'' of $\sim 5\times 10^{12}\hMsun$ that separates a population of low mass haloes with surrounding structure predominantly in the direction of their angular momentum, and another population of high mass haloes with higher correlation in the perpendicular direction. This has been previously reported by several authors, with different approaches and methodologies \citep{aragoncalvoetal2007,codisetal2015,pazetal2008}. Given that $\Mh$ haloes are, by construction, the third of the population that best fits the predictions in our TTT implementation, the relative agreement between this sample and the general population in Figure \ref{fig:gp_bines} suggests that {specific implementations} of the TTT, such as the anisotropic tidal torque theory \citep{codisetal2015}, may be adequate descriptions of the processes that build the angular momentum of $\Mh$ haloes. However, when we analyze the halo samples whose angular momentum growth deviates from our TTT implementation, no transition mass can be observed in any case. Although $\Wh$ haloes seem to follow a similar trend to that of the general population, their $\xi_\perp/\xi_\parallel$ quotients are systematically higher. In consequence, even at low masses, haloes of the $\Wh$ sample show a stronger correlation in the direction perpendicular to their angular momentum. The $\Lh$ sample, on the other hand, not only does not show a transition mass, but does not even follow the tendency of the general population. Rather, $\Lh$ haloes exhibit a mass independent trend of alignment between their angular momentum direction and the present-day mass distribution. 

These results could be indicating that $\Wh$ and $\Mh$ haloes experience analogous effects in their post-TTT angular momentum acquisition process, with these effects being strongly enhanced by the excess of structure in the plane perpendicular to the direction of rotation. 
{On the basis of Figure }\ref{fig:gp_bines}{, it is reasonable to expect a continuous behavior of the anisotropy-mass relation as a function of the halo angular momentum growth. In that case, our results could be indicating that the reported transition mass may be a feature that continuously depends on the deviations from the TTT: higher angular momentum growths imply lower transition masses. This analysis can be extended to the $\Lh$ sample, whose transition mass would be above $10\times^{14}\hMsun$.}
{The signal of $\Lh$ haloes toward low masses, on the other hand, could be related to an early loss of the initial surrounding tidal fields.} Given the highly clustered environments in which low mass $\Lh$ haloes inhabit (Figure \ref{fig:iso_M123}), it is likely to expect the emergence of external tidal fields whose main axes do not necessarily coincide with the principal directions of the initial surrounding matter distribution. This effect, together with the large amount of time that low mass haloes spent in the non-linear regime, could affect negatively their angular momentum acquired by tidal torquing, as well as diminish their rotational coherence.

\section{Summary and conclusions}
\label{conclusions}

In this paper we have studied how, from a classification based on systematic deviations of the angular momentum growth with respect to the TTT, we are able to define samples of DM haloes in numerical simulations with similar mass and angular momentum distributions, but with significantly different dynamical properties (Figures \ref{fig:gp_spinparameter} and \ref{fig:histo_Jxejes}) {and assembly history (Figure }\ref{fig:acrecion_M123}{)}. On the one hand, haloes that have won angular momentum above the TTT predictions ({winner} or $\Wh$) have particularly high rotational support and strong internal alignment, i.e. their angular momentum direction and shape axes are significantly correlated. {In addition, we found that $\Wh$ haloes have typically assembled the majority of their mass more recently, while showing a higher fraction of their mass distributed between their main progenitor and their second most massive progenitor}. On the other hand, haloes whose angular momentum growth is lower than what is expected from the model ({loser} or $\Lh$) show unusually low values of $\lambda$ {, a rather weak internal alignment and early formation times}. {Furthermore}, with the remaining population we have defined a third sample of haloes whose net angular momentum growth represents, by construction, the expected behavior in our TTT implementation, ({median} or $\Mh$). These haloes show dynamical properties consistent with the results obtained by other authors, such as the present-day spin parameter distribution.

By means of the isotropic halo-halo correlation function, we have statistically studied how the LSS is clustered around these samples, and how this trend depends on the mass of the haloes (Figure \ref{fig:iso_M123}). These functions are used to characterize the typical environments in which DM haloes form and acquire their angular momentum. We have found that low mass haloes tend to inhabit higher clustered regions if they belong to the $\Lh$ sample (from $\sim 30$ per cent to $\sim 50$ per cent above the general population of the same mass), but they are typically surrounded by a much less correlated structure if they belong to the $\Wh$ or the $\Mh$ samples (about $20$ per cent below the general population). On the other hand, although high mass haloes naturally form the densest structures of the Universe, there is a remarkable bias as to how these regions are occupied in regard of our classification. While high mass $\Wh$ haloes are mostly found in environments $\sim 20$ per cent more clustered than the general population of the same mass, $\Lh$ haloes inhabit regions with a lack of correlation of about $20$ per cent. High mass haloes of the $\Mh$ sample exhibit a correlation almost indistinguishable from that of the general population. {The overlapping of every correlation function below $\sim 2\hMpc$ for high mass haloes suggests that any influence of the clustering on the systematic deviations from the TTT has to be due to the LSS rather than the immediately surrounding matter distribution.}

Through an analysis of anisotropic correlation functions, we have determined differences on the orientation of the structure around haloes from each sample (Figure \ref{fig:ani_M123}). 
At low masses, we have found that $\Lh$ haloes show in both directions, perpendicular and parallel to their angular momentum, a significantly higher correlation with the surrounding structure than the other two samples, with a slight tendency of this relative excess to decrease with distance. There are, however, considerable differences in the orientation of the structure that surrounds haloes of the $\Wh$ and $\Mh$ samples. Whereas along the axis of rotation both samples show indistinguishable curves of correlation within the entire distance range, near the plane perpendicular to their angular momentum $\Wh$ haloes are significantly more correlated with their neighbors than $\Mh$ haloes.
At high masses the anisotropies are more noticeable. None of the parallel correlation functions show appreciable differences with respect to the general population, thus the matter distribution along the direction of the angular momentum of high mass haloes seems to have weak or negligible influence on their deviations from the TTT. The signals observed in the isotropic analysis, i.e. the over-clustering around $\Wh$ haloes and the relative lack of neighbors around $\Lh$ haloes, are due exclusively to the correlation with the structure near the plane perpendicular to the angular momentum. In other words, the preferred direction of the rotational axis of $\Wh$ haloes is perpendicularly oriented with respect to the dominant structure, and the lacking of this alignment, which gradually reduces toward $\Mh$ and $\Lh$ haloes, seems to cause high mass haloes to decrease their angular momentum growth.

Finally, we studied the preponderance of parallel or perpendicular structure around each sample. Analyzing mass-dependent quotients between the anisotropic correlations (Figure \ref{fig:gp_bines}), we were able to reproduce, for the general population, the well known correlation between the mass and the perpendicularity of the predominant structure with respect to $\mathbfit{J}$. Moreover, we found that the halo sample that best fits the predictions in our TTT implementation, i.e. the $\Mh$ sample, shows the best agreement with previous results of this relation. 
However, we found a novel and strong dependence of this trend on the systematic deviations from the TTT. Whereas $\Wh$ haloes follow the same tendency that the general population with a systematically higher signal in the perpendicular direction, $\Lh$ haloes show a rather mass-independent behavior, with their surrounding structure predominantly aligned with their angular momentum. 
{The continuity that appears to be from $\Lh$ haloes to $\Wh$ haloes through the $\Mh$ sample suggests that the reported transition mass }\citep{aragoncalvoetal2007,pazetal2008,codisetal2015}{, could be a feature that continuously depends on the angular momentum growth. This hypothesis should be further tested in simulations with a wider dynamical range or with a larger number of DM haloes in order to increase the statistical significance.}

The fact that the $\Mh$ sample reproduces previous results, such as the present-day distribution of the spin parameter and the transition mass of $\sim 5\times10^{12}\hMsun$, allows us to hypothesize that haloes that show an angular momentum growth consistent with the predictions of the TTT represent a part of the population less disturbed during their formation and, consequently, with more standard and predictable properties.
Conversely, our results suggest that the differences between $\Wh$ and $\Lh$ haloes (i.e. haloes that deviate from the TTT) can be associated to the characteristics of the environment in which they form and with which they interact, especially during the non-linear stages. 

Regardless of mass, $\Wh$ and $\Mh$ haloes show the same clustering in the direction of the angular momentum, but near its perpendicular plane $\Wh$ haloes are typically surrounded by a more correlated structure. Moreover, the quotient between the perpendicular and parallel correlations of the $\Wh$ sample, although it is higher, has the same mass dependence as the $\Mh$ sample.
These correspondences could be indicating that $\Wh$ and $\Mh$ haloes are affected by analogous processes, but differences could arise during the non-linear regime, in a sort of secondary tidal torque, if the anisotropy of the surrounding tidal fields continue to favor the angular momentum growth of $\Wh$ haloes. {This scenario seems to be supported by the results obtained for the assembly history, which show that at recent times $\Wh$ haloes are more profusely accreting material and their mass is more fragmented. This allow us to expect a delaying in the decoupling of their inertia tensor from the influence of the neighboring structures, hence higher angular momentum growths by tidal torquing and, consequently, more coherent spin acquisition. For high masses, consistently with} \citet{faltenbacherywhite2010}{, there is a correlation between enhanced clustering and higher spins. Moreover, our results complement the previous analysis, showing that more rotationally supported haloes not only have more neighbors, but also have them close to the plane perpendicular to their angular momentum, suggesting the existence of effective tidal torques at recent times.}

Haloes of the $\Lh$ sample show systematically more correlation with the structure in the direction of their angular momentum. The poor rotational support and low level of alignment between their shape and direction of rotation could be indicating that the environment of these haloes has typically suffered disturbances that did not preserve the coherence of the initial surrounding tidal fields. {Interestingly, the alignment between the angular momentum and the dominant surrounding structure found in the low mass range, in addition to the early accretion and enhanced clustering, are in excellent agreement with the results from} \citet{borzyszkowskietal2017}. 
Their ``stalled'' haloes (i.e. early-collapsing haloes that stopped accreting mass due to the presence of strong tides in filaments) seem to fit remarkably good with our $\Lh$ sample, which suggests not only a strong correlation between the assembly bias and the deviations from the TTT, but also the possibility that $\Lh$ haloes are typically found within thick filaments. This is consistent with our results and allow us to conjecture that the gravitational influence of thick filaments on infalling particles and, consequently, in the accretion flow of $\Lh$ haloes, could play a substantial role in the angular momentum acquisition, as described by \citeauthor{borzyszkowskietal2017}. {Moreover, we found a statistically significant tendency of $\Lh$ haloes to have low $\lambda$ values. Although }\citeauthor{borzyszkowskietal2017}{ do not include the result in their conclusions due to the low number of their sample, it is to expect that stalled haloes also show low rotational support.}
However, even though the environment that surrounds $\Lh$ haloes presents anisotropies that do not depend on their mass, these haloes show different clustering levels with respect to the general population at different masses. This suggests that the deviations from TTT observed at high and low masses could be produced by different mechanisms, and motivate us to continue studying their environment{ and accretion history}.

In a future work we will deepen the analysis of the typical regions that haloes with different angular momentum growth inhabit. Specifically, we will identify the topology of the halo surrounding structures (voids, walls and filaments). These new efforts will be aimed at unraveling the possible mechanisms responsible for the correlations found in this work.


\section*{Acknowledgements}
We thank M. A. Sgr\'o and F. Stasyszyn for useful comments on this work. We would also like to thank the anonymous referee for the helpful suggestions and corrections. PL acknowledges the receipt of a fellowship granted by the Argentine \emph{Agencia Nacional de Promoci\'on Cient\'ifica y Tecnol\'ogica} (ANPCyT), under reference PICT 2015-03098. 





\bibliographystyle{mnras}
\bibliography{bibliografia} 








\bsp	
\label{lastpage}
\end{document}